\documentclass[a4paper]{jpconf}
\usepackage{graphicx}
\usepackage{amstext,amsmath,amssymb,amsfonts,bbm}
\def\be{\begin{equation}}
\def\ee{\end{equation}}
\def\bes{\begin{eqnarray}}
\def\ees{\end{eqnarray}}

\def\6{\langle}
\def\9{\rangle}

\def\1{{\mathbbm 1}}

\begin{document}
\title{Divergence on the horizon}

\author{Judy Kupferman}

\address{Ben Gurion University, Beer Sheva, Israel
}

\ead{Judithku@bgu.ac.il}

\begin{abstract}
Black hole entropy has been shown by 't Hooft to diverge at the horizon,
whereas entanglement entropy in general does not. We show that because
the region near the horizon is a thermal state, entropy is linear
to energy, and energy at a barrier is inversely proportional to barrier
slope, and diverges at an infinitely sharp barrier as a result of
position/momentum uncertainty. We show that 't Hooft's divergence at
the black hole is also an example of momentum/position uncertainty,
as seen by the fact that the {}``brick wall'' which corrects it
in fact smooths the sharp boundary into a more gradual slope. This
removes a major obstacle to identification of black hole entropy with
entanglement entropy.

\end{abstract}

\section{Introduction}

Two of the possible characterizations of black hole entropy are both
proportional to area but they apparently do not coincide. One is entanglement
entropy, arising from correlations between that within the black hole
and that without \cite{Srednicki}\cite {CallanWilczek}\cite{SusskindUglum}, and
the other is the statistical mechanics definition related to the number
of states of a system in the vicinity of a black hole \cite{GibbonsHawking}\cite{t'Hooft}.
Both types of entropy are expressed by calculating $-Tr(\rho ln\rho)$
of fields exterior to the black hole and the question naturally arises:
do they both refer to the same thing? One problem in uniting these
descriptions is the question of divergence at the black hole horizon.
Entanglement entropy may have UV divergence which will need to be renormalized,
but it does not diverge at a particular location. However, in the
statistical mechanics calculation first carried out by 't Hooft, the
integrand diverges at the black hole boundary. We will show that the
problem of divergence at the boundary can be resolved. It is not a
unique black hole characteristic but rather a result of quantum uncertainty,
and the correct expression must involve smearing out the boundary.
In this way the two expressions may coincide, and we may consider
whether black hole entropy is simply entanglement entropy. This talk
is based on \cite{ours}.

For a massive black hole in the region just outside the horizon, it
has been shown that entanglement entropy coincides with the thermodynamic
definition of entropy \cite{KabatStrassler}\cite{RamyAmosMarty}. Thermodynamic
entropy is linear to energy. The expectation value for energy is also
linear to momentum uncertainty, $\left\langle E\right\rangle \sim\left\langle p^{2}\right\rangle =\Delta p^{2}$.
Therefore whenever energy is calculated at a precisely defined location
we expect it to diverge due to the uncertainty relationship between
$x$ and $p$ . We show that when any system is cut off by a barrier,
the increase in energy is proportional to the narrowness of the barrier,
and diverges as the barrier becomes infinitely sharp. Thus black hole
entropy, which is proportional to energy, will diverge on the horizon
because the horizon is a sharply defined barrier. We also show that
this is consistent with 't Hooft's {}``brick wall'' cutoff.

First we review these two characterizations of black hole entropy,
entanglement and statistical, focusing on behavior at the horizon.
Then we show in detail the relationship between energy and entropy
in a massive black hole. We examine behavior of energy at a boundary
between two subsystems, first for the non-relativistic and then for
the relativistic case, and show that in both cases energy diverges
as the boundary becomes sharp. Finally, we examine 't Hooft's calculation
leading to divergence at the boundary, and find that his relocation
of the boundary to avoid divergence is equivalent to smearing out
the boundary. Therefore here too the divergence is related to sharpness
of the boundary, and thus it reflects behavior of any thermal system
at a boundary, and is not unique to a black hole. This conclusion
lends weight to the possibility that black hole entropy is in fact
entanglement entropy, because the peculiar divergence which seemed
to single it out from other entangled systems is just a result of
quantum x/p uncertainty.

\section{Entropy at the horizon}

The area law was first formulated by Bekenstein, beginning from the
observation that black hole area must increase as does entropy \cite{Bekenstein},
and strengthened by thermodynamic calculations \cite{GibbonsHawking}. Bekenstein did not call this entanglement entropy
but he did describe it as a measure of the inaccessibility of information
to an exterior observer as to which internal configuration of the
black hole is realized in a given case. In fact black holes have entanglement
entropy by definition. Entanglement entropy quantifies the extent
to which a state is mixed. If the universe is described as a pure
state, the black hole horizon divides it into that within and that
without the hole, each of which is a mixed state. Therefore black
holes have entanglement entropy, and the question is whether black
hole entropy is anything more, or whether entanglement entropy saturates
the definition.

't Hooft calculated thermodynamic characteristics of a black hole,
among them entropy, and in doing so found a divergence of the integrand
at the horizon \cite{t'Hooft}. He overcame the problem by adjusting the limits
of integration to a {}``brick wall'' a finite infinitesimal distance
from the horizon, and various other solutions have been adopted since
\cite{Myers}. Entanglement entropy, on the other hand,
has ultraviolet divergence, and a UV cutoff must be employed, but
it does not diverge at any particular location. Srednicki in his landmark
paper \cite{Srednicki} calculated entropy by tracing over the degrees of freedom
of part of the system, and found it proportional to area as well.
He numerically obtained an expression for entropy $S=0.30M^{2}R^{2}$
,where $R=(n+\frac{1}{2})a,$ $a$ is lattice spacing and $n$ the
number of discrete oscillators. He also defines $M$ as the inverse
lattice spacing $a^{-1}$ so that the actual expression is $S=0.30(n+\frac{1}{2})^{2}$
which diverges for an infinite number of oscillators, but not at a
particular location. Other treatments of entanglement entropy in general
\cite{Bombelli}\cite{Plenio} also find UV divergence but not at any particular
location. Therefore horizon divergence would seem be a special characteristic
of black hole entropy, and would preclude its being a manifestation
of entanglement entropy as such. We show below that this is not the
case.

For a massive BH in equilibrium the space just outside the hole near
the horizon can be treated as a thermal state in Rindler space \cite{KabatStrassler}\cite{RamyAmosMarty}.
In this case entanglement entropy coincides with thermal
entropy, as follows. To find entanglement entropy we take the trace
of part of the system. If that part of the system is a thermal state,
the partial trace is a thermal density matrix. {\small \begin{eqnarray}
S & = & -Tr\left(\rho_{part}ln\rho_{part}\right)\nonumber \\
\rho_{part} & = & \frac{1}{Z}\sum_{i}e^{-\beta E_{i}}|E_{i}><E_{i}|\nonumber \\
\left\langle E\right\rangle  & = & \frac{1}{Z}\sum_{i}E_{i}e^{-\beta E_{i}}\nonumber \\
S & = & (-\frac{1}{Z}\sum_{i}e^{-\beta E_{i}})\times(-\beta\sum_{i}E_{i}-lnZ)\nonumber \\
 & = & \beta<E>+lnZ.\end{eqnarray}
}For a scalar field with finite temperature ln Z is a constant, so
the entropy is linear to the expectation value of the energy. Therefore
in the case of a massive black hole the entanglement entropy behaves
as does the energy. So instead of calculating entropy, which is a
complicated non local quantity, we can calculate the reduced density
matrix of a subsystem and look at the behavior of its energy, which
is a simpler local quantity.

\section{Nonrelativistic treatment}

 We first write down the expectation value for energy in a nonrelativistic
many body system. In order to look at behavior at a boundary, we insert
a window function into the system. The window function mimics the
horizon by {}``truncating'' part of space for the particles. However
it does not affect boundary values of the system. An example of such
a window function would be the Heaviside step function. Here we provide
our window function with a varying slope to see how sharp localization
affects the energy divergence.

In a nonrelativistic many body system, taking spinless particles for
simplicity, the field state and the kinetic energy operator for a
free particle are as follows;\begin{eqnarray*}
\left|\Psi^{\dagger}(\vec{r})\right\rangle  & = & \Psi^{\dagger}(\vec{r})\left|0\right\rangle =\sum_{\vec{p}}\frac{e^{-i\vec{p}\vec{r}}}{\sqrt{V}}\, a_{\vec{p}}^{\dagger}\left|0\right\rangle \end{eqnarray*}
The energy operator is $T=\sum_{\vec{p}}\frac{p^{2}}{2m}a_{\vec{p}}^{\dagger}a_{\vec{p}},$
and its expectation value is $\left\langle E\right\rangle =\left\langle 0|\Psi T\Psi^{\dagger}|0\right\rangle $.
In configuration space \begin{equation}
\Psi T\Psi^{\dagger}=\intop_{-\infty}^{\infty}d^{3}r\frac{1}{2m}\nabla_{r}\Psi\left(\vec{r}\right)\nabla_{r}\Psi^{\dagger}\left(\vec{r}\right)\end{equation}
With a window function $f(\vec{r})$, the field operator becomes \[
\Psi_{W}=\int d^{3}r\, f\left(\vec{r}\right)\Psi^{\dagger}\left(\vec{r}\right)=\int d^{3}r\, f\left(\vec{r}\right)\sum_{\vec{p}}\frac{e^{-i\vec{p}\vec{r}}}{\sqrt{V}}\, a_{\vec{p}}^{\dagger}\]
The ladder operators on the vacuum give delta functions, resulting
in \begin{eqnarray}
\left\langle E\right\rangle  & = & \frac{1}{2m}\intop_{-\infty}^{\infty}d^{3}r\nabla_{r}f^{*}\left(\vec{r}\right)\nabla_{r}f(\vec{r})\label{eq:nonrelativistic_energy_eqs-2}\end{eqnarray}
For details see \cite{ours}. This represents the contribution
of the window to the expectation value of the energy.%
\footnote{In addition the state would have a wave function, e.g. $g(\vec{r})$,
which would give us the energy of the particle regardless of window,
but here we have taken $g(\vec{r})=1$ for simplicity.%
}

\begin{figure}[h]
\centering\includegraphics[scale=0.6]{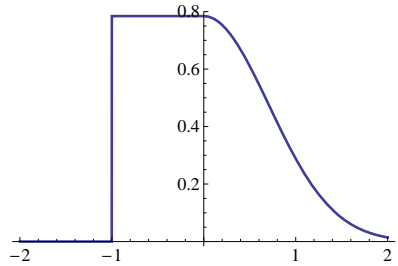}\caption{\label{fig:half gaussian}Half-Gaussian window, bounding the wave
function, where we look at the part on the left of the slope. The
horizon varies from a sharp step at x=0 to a gradual slope as in the
drawing. The step at $x=-1$ is just for normalization; in fact space
on the left continues to $-\infty$. }

\end{figure}

We take the window function to represent a boundary with a varying
slope. Technically we achieve this by combining a Gaussian with the
Heaviside step function, and bound the step for purposes of normalization
(see figure \ref{fig:half gaussian}). In the figure, the amplitude
to see the particle through the window falls off gradually with the
slope, so that on the left side (from $-1$ to $0$ in the graph)
it is constant, and then it slopes down to zero with width L. The
sharp drop on the left (at $-1$ in the graph) is for purposes of
normalization, but does not affect the result; taking the cutoff at
other values gives essentially the same result: energy which is inversely
proportional to L . This is graphed in one dimension for clarity of
graphing, but three dimensions give a symmetrical version of the same
results.

 We calculate the contribution to the energy expectation value for
a system with this window function and obtain

\begin{figure}[h]
\centering \includegraphics[scale=0.8]{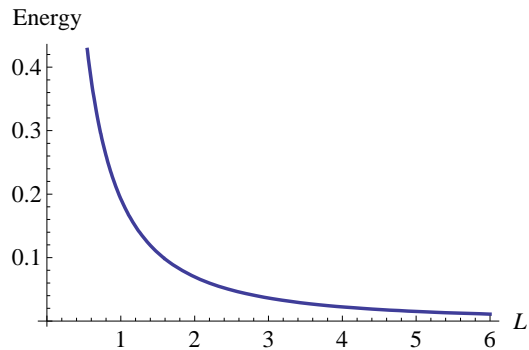}
\caption{\label{fig:-deltah_for_halfgauss}$\left\langle E\right\rangle $
for sloping window, as a function of slope width.}

\end{figure}

{ \begin{eqnarray}
\left\langle E\right\rangle  & = & \frac{1}{2m\left(L^{2}+2L\sqrt{\frac{2}{\pi}}\right)}\end{eqnarray}
}where $L$ refers to width of the slope and $m$ is particle mass.
Obviously this diverges when the slope becomes perfectly sharp, L=0,
as seen in figure \ref{fig:-deltah_for_halfgauss} where we have taken
$m=1$.

We have obtained the same result using completely different functions
for the slope. For example, taking the window function as arctangent
gives a slope which can be varied by using $Arctan[nx]$; larger n
gives steeper slope. We took $\sqrt{Arctan[nx]+\frac{\pi}{2}}$ so
that all values would be positive and to simplify the numerical calculations,
and again the energy was seen to be inversely proportional to width
of slope. Thus we have seen that the energy increases as the barrier
grows sharper. That is, the more sharply the position of the dividing
barrier is specified, the closer the energy gets to divergence. We
now note that $E=\frac{1}{2m}\left\langle \psi\right|p^{2}\left|\psi\right\rangle $.
Since $\left\langle \psi\right|p\left|\psi\right\rangle =0$, $\left\langle \psi\right|p^{2}\left|\psi\right\rangle =\left(\Delta p\right)^{2}$,
so that as energy diverges so do fluctuations in momentum. This divergence
occurs as the width is taken to zero, and may be seen as an example
of position/momentum uncertainty.

\section{Relativistic treatment}

In a relativistic system the energy operator is taken from the energy
momentum tensor: $\hat{H}=T^{00}=\int\frac{d^{3}k}{\left(2\pi\right)^{3}}k_{0}a_{k}^{\dagger}a_{k}$
taking $c=\hbar=1$. A state with window function, as before, is $\int d^{3}r\, f\left(\vec{r}\right)\Psi^{\dagger}\left(\vec{r}\right)\left|0\right\rangle $
where the field operator here is the relativistic one.

In order to look for the various expectation values we need the relativistic
scalar product:\begin{equation}
\left\langle \varphi|\phi\right\rangle =\frac{1}{2}\int d^{3}x\left(\varphi i\partial_{0}\phi-\left(i\partial_{0}\:\varphi\right)\phi\right)\end{equation}
where the expression is sandwiched between the vacuum.%
\footnote{If this use of scalar product is unfamiliar, note that $\left\langle \widehat{H}\right\rangle =\left\langle \varphi\left|i\partial_{0}\right|\varphi\right\rangle $
with this formula gives the same expression as $\left\langle \int d^{3}rT^{00}\right\rangle $
. %
} Thus for energy we write

\begin{equation}
\left\langle \varphi\left|\widehat{H}\right|\varphi\right\rangle =\frac{1}{2}\int d^{3}x\left(\varphi i\partial_{0}\left(\widehat{H}\varphi\right)-\left(i\partial_{0}\:\varphi\right)\widehat{H}\varphi\right).\end{equation}
Next we plug in the state with the window function. Since in the relativistic
case $E^{2}\sim p^{2}$ and we are interested in momentum fluctuations,
we calculate $\left\langle \varphi\left|\widehat{H}^{2}\right|\varphi\right\rangle $
and obtain \begin{equation}
\frac{1}{2}\intop_{-\infty}^{\infty}d^{3}r\nabla_{r}f^{*}\left(\vec{r}\right)\nabla_{r}f(\vec{r}).\label{eq:relativistic energy expec value}\end{equation}
For details see \cite{ours}. This clearly has the same behavior as the non relativistic
case.

We have seen that in just as in the nonrelativistic treatment, energy
is seen to diverge the more sharply position is specified. For the
non relativistic case we had $E\sim p^{2}$ whereas in the relativistic
case $E^{2}\sim p^{2}$ but in both cases the divergence may be seen
as an example of position/momentum uncertainty.

\section{Statistical mechanics calculation}

't Hooft treats the black hole as a quantum system, and solves the
wave equation in the Schwartzschild metric:\begin{equation}
\left(1-\frac{2M}{r}\right)^{-1}E^{2}\varphi+\frac{1}{r^{2}}\partial_{r}\left(r\left(r-2M\right)\partial_{r}\right)\varphi-\left(\frac{l\left(l+1\right)}{r^{2}}+m^{2}\right)\varphi=0.\label{eq:thooft wave eq}\end{equation}
 He identifies $k$, the wave number, by isolating the radial derivative
on one side of the equation and the rest of the expression on the
other. Thus he obtains an expression for $k^{2}(r)\varphi$, where
$k^{2}(r)$ is the eigenvalue of the second radial derivative. Labelling
the metric coordinates by indices for short, he obtains\[
k^{2}=g_{rr}\left(g^{tt}E^{2}-\left(\frac{l\left(l+1\right)}{r^{2}}+m^{2}\right)\right)\]
Using a WKB approximation he obtains the number of modes by integrating
the wave number over the region outside the black hole horizon:\begin{eqnarray}
\pi N & = & \intop_{2M}^{L}dr\, k\left(r,l,m\right)\label{eq:t'Hooft integral}\\
 & = & \intop_{2M}^{L}dr\left(1-\frac{2M}{r}\right)^{-1}\int(2l+1)dl\sqrt{E^{2}-\left(1-\frac{2M}{r}\right)\left(\frac{l\left(l+1\right)}{r^{2}}+m^{2}\right)}\nonumber \end{eqnarray}
where L is an infrared cutoff. However the redshift leads this to
diverge at the lower limit, that is, the horizon itself. To avoid
this 't Hooft takes the lower limit a slight distance away from the
horizon, his well known {}``brick wall,'' so that the lower limit
becomes $2M+h$ . From this expression he obtains the energy and entropy,
which diverge as $h\rightarrow0$.

\begin{figure}[h]
\centering \includegraphics[scale=.8]{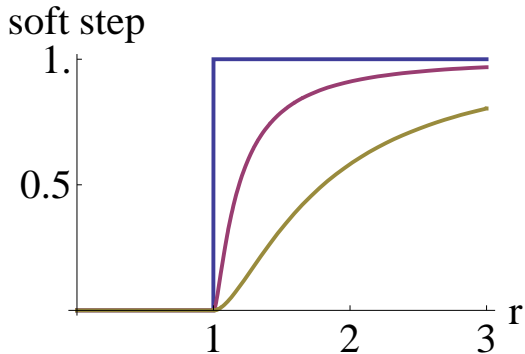}
\caption{$\tilde{\Theta}(r-2M)$ at the\label{Flo:softstep} horizon as function of $r/2M$. Curves have $h=0$  (sharp step), $h=0.1$ and $h=0.9$ (lowest) }
\end{figure}

't Hooft's adjustment of the lower bound of the integral from $2M$
to $2M+h$ is equivalent to a change of variable which leaves the
bound at $2M$ but adds a factor into the redshift: \begin{eqnarray*}
\intop_{2M+h}^{L}dr\left(1-\frac{2M}{r}\right)^{-1} & = & \intop_{2M}^{L}d\tilde{r}\left(1-\frac{2M}{\tilde{r}+h}\right)^{-1}\end{eqnarray*}
We examine the change in the redshift. In the $\tilde{r}$ system
the redshift is \[
\left(1-\frac{2M}{\tilde{r}}\right)^{-1}.\]
So the altered redshift is the same as taking the redshift and multiplying
it by another function, thus:

\begin{eqnarray}
\left(1-\frac{2M}{\tilde{r}+h}\right)^{-1} & = & \left(1-\frac{2M}{\tilde{r}}\right)^{-1}\: f(r,h)\nonumber \\
f\left(r,h\right) & = & \frac{\left(\tilde{r}+h\right)\left(\tilde{r}-2M\right)}{\tilde{r}\left(\tilde{r}-2M+h\right)}\nonumber \\
\intop_{0}^{L}drg_{rr}(r)f(r,h)\Theta(r-2M) & = & \intop_{0}^{L}drg_{rr}(r)\tilde{\Theta}(r-2M)\label{eq:soft step}\end{eqnarray}
 Thus we see that 't Hooft's changed lower limit is exactly equivalent
to modification of the step function. The new step function is no
longer a sharp step, but rather has a width of $h$ (for $r=2M+h$,
$\tilde{\Theta}=1/2$). Figure \ref{Flo:softstep} shows a graph of
this softened step function $\tilde{\Theta}(r-2M)$ for various values
of $h$, ranging from a sharp step at $h=0$ to a wider and wider
slope. For $h=0$ 't Hooft had a divergence and when he changed the
lower limit, equivalent to widening the horizon slope, the divergence
vanished.

This is in fact position/momentum uncertainty. A closer look at eq.\ref{eq:t'Hooft integral}
and at the effect of the modifying function $f(r,h)$ makes this clear.
The integrand of eq.\ref{eq:t'Hooft integral} is $k(r)$ . Just as
in the previous sections $\left\langle k^{2}\right\rangle =\Delta k^{2}$
, and its root is just the integrand of eq. \ref{eq:t'Hooft integral}.
Multiplying this integrand by the modifying function $f(r,h)$ is
equivalent to multiplying it by $\Delta x$ since as we have shown,
it widens the slope. Eq. \ref{eq:t'Hooft integral} with the {}``brick
wall'' expressed as a softening function becomes $\intop_{2M}^{L}dr\Delta k\Delta x$.
Without the brick wall, $\Delta x=0$ and we had $\Delta k\rightarrow\infty$.
Adding the brick wall is directly equivalent to widening $\Delta x$
and as a result the expression becomes finite.

Another way of looking at it is that the {}``brick wall'' is equivalent
to a modification of the redshift: $g_{rr}\rightarrow\tilde{g}_{rr}=g_{rr}f(r,h)$.
This may become more intuitive from the following illustration: I
hover outside a black hole and send an astronaut in towards it radially,
and require of the astronaut to send me back progress reports every
kilometer; we all know the progress reports will come more and more
slowly, so that it seems to me that he has slowed down, and come to
a full stop at the horizon. I send another astronaut after him and
the same thing happens. So I send a whole fleet of astronauts, tied
together by a rope at regular intervals, as in dangerous mountain
climbing. It will seem to me that as they near the horizon they all
slow down, the rope slackens so they grow closer and closer, and at
the horizon eventually they will all crowd together and I'll have
an infinite density of astronauts there. This is the effect of the
redshift. However, if I smooth out the horizon -  stretch out the
redshift, so to speak - they will not crowd up at one point and the
density won't be infinite. This is, so to speak, quantum uncertainty
delta(x) delta(astronaut).

't Hooft's displacement of the boundary prevents divergence at the
black hole horizon by distorting the horizon so that it is no longer
a sharply located step function but rather a more gradual transition.
By doing so he actually shrinks the momentum uncertainty. Therefore
we see that this too is an example of $x/p$ quantum uncertainty,
and not a unique characteristic of a black hole.

\section{Discussion}

Energy has been shown to diverge as the boundary between two subsystems
becomes sharp. The divergence is due to the fact that the energy is
a simple function of momentum. For the nonrelativistic case it is
easy to see that $\left\langle E\right\rangle =\frac{1}{2m}\left\langle p^{2}\right\rangle =\frac{1}{2m}\Delta p^{2}$and
so divergence at a sharp boundary is just due to the quantum $\Delta p\Delta x$
uncertainty. The relativistic expression for $\left\langle E^{2}\right\rangle $has
exactly the same form as the nonrelativistic result and so in both
cases energy divergence at an infinitely sharp boundary is a consequence
of x/p uncertainty.

The region near the boundary of a black hole is a thermal state, where
the entropy is linear to energy. Therefore black hole entropy will
diverge at the boundary as well. We have not proven that there is
no other cause of the divergence, unique to a black hole. But we have
shown that regardless of any other cause, there would be divergence
at the boundary as a result of the uncertainty principle. We have
also shown that 't Hooft's divergence at the black hole is also an
example of momentum/position uncertainty, as seen by the fact that
the {}``brick wall'' which corrects it in fact smooths the sharp
boundary into a more gradual slope.

We may now consider whether the entanglement and statistical mechanics
definitions of black hole entropy might refer to the same thing. Both
are proportional to area. The UV divergence may be renormalized with
a cutoff, and the boundary divergence by smearing out the boundary,
so that these no longer preclude unification of the two expressions.
The argument can then be made that black hole entropy is due to entanglement,
that is, to quantum correlations between the two parts of the system
inside and outside the black hole, and that counting the number of
states is tantamount to counting the correlations.

However there is a third definition with which these two must now
be reconciled. Black hole entropy has also been shown - from thermodynamic
considerations \cite{Bekenstein} as well as explicit calculations in string
theory \cite{StromingerVafa} to equal one fourth of the horizon area. An open problem
is to obtain the factor of $1/4$ in these two cases as well.
\\

 This work has been supported by grant
239/10 of the Israel Science Foundation. We would like to thank Merav Hadad for many useful discussions.

%------------------------------------------------------------------------------------------------------------------------

\section*{References}

\bibliographystyle{unstr}

\end{document}